\documentclass[12pt, preprint]{aastex}
\shorttitle{}
\shortauthors{Nesvorn\'y et al.}

\usepackage{amsmath}
\usepackage{lineno}
\newcommand{\bmath}[1]{\mbox{\boldmath{$#1$}}}

\newcommand{\cs}{c_{\rm s}}

\begin{document}
\baselineskip 19.pt

\title{Trans-Neptunian Binaries as Evidence for Planetesimal Formation by the Streaming Instability}
\author{David Nesvorn\'y$^{1,*}$, Rixin Li$^2$, Andrew N. Youdin$^2$, Jacob B. Simon$^{1,3}$,
William M. Grundy$^4$}
%, Keith S. Noll$^5$, Derek C. Richardson$^6$}
\affil{$^1$Department of Space Studies, Southwest Research Institute,
1050 Walnut St., Suite 300, Boulder, CO 80302, USA}
\affil{$^2$Steward Observatory \& Department of Astronomy, University of Arizona, 933 N. Cherry Avenue, 
Tucson, AZ, 85721, USA}
\affil{$^3$JILA, University of Colorado, 440 UCB, Boulder, CO 80309, USA}
\affil{$^4$Lowell Observatory, 1400 W. Mars Hill Rd., Flagstaff, AZ 86001, USA}
\affil{*e-mail:davidn@boulder.swri.edu}
\maketitle

{\bf A critical step toward the emergence of planets in a protoplanetary disk consists 
in accretion of planetesimals, bodies 1-1000 km in size, from smaller disk constituents. 
This process is poorly understood partly because we lack good observational constraints on 
the complex physical processes that contribute to planetesimal formation [1]. In the outer solar system, 
the best place to look for clues is the Kuiper belt, where icy planetesimals survived to 
this day. Here we report evidence that Kuiper belt planetesimals formed by the streaming 
instability, a process in which aerodynamically concentrated clumps of pebbles gravitationally 
collapse into $\sim$100-km-class bodies [2]. Gravitational collapse was previously suggested 
to explain the ubiquity of equal-size binaries in the Kuiper belt [3,4,5]. We analyze 
new hydrodynamical simulations of the streaming instability to determine the model expectations 
for the spatial orientation of binary orbits. The predicted broad inclination distribution 
with $\simeq$80\% of prograde binary orbits matches the observations of trans-Neptunian
binaries [6]. The formation models which imply predominantly retrograde binary orbits (e.g., [7]) 
can be ruled out. Given its applicability over a broad range of protoplanetary disk 
conditions [8], it is expected that the streaming instability seeded planetesimal formation 
also elsewhere in the solar system, and beyond.}  

The streaming instability (SI) is a mechanism to seed planetesimal formation by aerodynamically 
concentrating particles to high densities [2,9-11]. SI simulations show that particle 
concentration is strong if the (local and height-integrated) solid-to-gas ratio is at least 
modestly enhanced over solar abundances [10]. Global disk 
evolution, including photoevaporation, ice lines, pressure traps and other effects (e.g., [12]) 
can readily produce the required enhancement, suggesting that the SI should commonly operate 
in protoplanetary disks to hatch planetesimals. Alternately, if the first planetesimals formed 
with maximum sizes of $\sim$1-10 km (by any mechanism), they can subsequently grow by accreting 
mass in mutual collisions, a gradual process known as the collisional coagulation (e.g., [13]).
Previous attempts to discriminate between different formation processes from the size distribution 
of planetesimals have been inconclusive [14,15].

We analyze a suite of vertically stratified 3D simulations of the streaming instability (SI) [16,17]. 
The simulations were performed with the {\tt ATHENA} code [18], which accounts for the hydrodynamic 
flow of gas, aerodynamic forces on particles, backreaction of particles on the gas flow, and particle
self-gravity. We used the shearing box approximation with at least 512$^3$ gas cells, more than 
$1.3\times10^8$ particles and appropriate boundary conditions (Methods). Each simulation was 
parametrized by the dimensionless stopping time of participating particles, $\tau=t_{\rm stop} \Omega$, 
where $\Omega$ is the Keplerian frequency, and the local particle-to-gas column density ratio, $Z$ 
(additional parameters are discussed in Methods). We adopted $\tau=0.3$-2, which would correspond 
to sub-cm-size pebbles in the Minimum Mass Solar Nebula (MMSN; [19]) at 45 au if the gas density 
was reduced by photoevaporation [12], and $Z=0.02$-0.1. Other choices of these 
parameters yield similar results [16,17] as long as the system remains in the SI regime [8].

As the time progresses in our simulations (Figure 1), dense azimuthal filaments 
form, fragment and condense into hundreds of gravitationally-bound clumps. We used an efficient tree-based 
algorithm ({\tt PLAN}; Methods) to identify all clumps (Figure 1c). 
Unfortunately, the resolution in the {\tt ATHENA} code does not allow us to follow the gravitational 
collapse of each clump into completion. 
Instead, we measure the total angular momentum, $J$, and its $z$-component $J_z = J \cos \theta$, 
giving the clump obliquity $\theta$.
The total angular momentum can be compared to that of a critically 
rotating Jacobi ellipsoid: $J_{\rm c}=0.39 (G M^3 r)^{1/2}$ [20], where $G$, $M$ and $r$ are the gravitational 
constant, mass and effective radius (obtained from $M$ with $\rho=1$ g 
cm$^{-3}$). We find that $J/J_{\rm c} \gg 1$ (this conclusion is insensitive to the choice of density), 
thus demonstrating that either most of the initial angular momentum must be removed or a typical SI 
clump cannot collapse into an {\it isolated} planetesimal.  

The vigorous rotation of the SI clumps established here is conducive to the formation of {\it binary} 
planetesimals with properties that closely match observations of the trans-Neptunian binaries.  Specifically, 
in the regime of $J/J_{\rm c} \gg 1$, gravitational collapse is capable of producing a $\sim$100\% binary 
fraction [3] consistent with observations [5]. Binary planetesimals produced by gravitational collapse 
have nearly equal-size components ($r_2/r_1>0.5$, where $r_1$ and $r_2$ are the primary and secondary 
component radii; Figure 2) and large separations ($a_{\rm b}/(r_1+r_2)>10$, where $a_{\rm b}$ is the binary 
semimajor axis), just as needed to explain observations [4-6]. Moreover, the matching colors of binary components 
[21] imply that each binary formed with a uniform compositional mix, as expected for gravitational 
collapse (but not random capture). Here we elaborate the prediction of the SI model for the spatial 
orientation of binary orbits.

The clump obliquities obtained in our SI simulations have a broad distribution (Figure~3) with 
$\simeq$80\% of clumps having $\theta<90^\circ$ (prograde rotation relative to the heliocentric orbit) 
and $\simeq$20\% having $\theta>90^\circ$ (retrograde rotation). This result is relatively insensitive 
to various SI parameters (e.g., $\tau$ and $Z$; Methods and Supplementary Figures 1-4) in the regime 
of strong particle clumping that has been explored. Other concentration mechanisms -- e.g., isotropic 
turbulence or secular gravitational instability -- lack concrete predictions for rotation, but are not 
expected to produce the same distribution of angular momenta by chance. The obliquity distribution 
shown in Figure 3 is thus a telltale signature of the SI.

To understand how clump obliquities were established, we traced their evolution back in time.
Two stages were identified. To get insights into the initial aerodynamic stage, we computed the 
vorticity of the particle field, $\nabla \times v_{\rm p}$, where $v_{\rm p}$ is the particle 
velocity. We found that the colatitude distribution of the vorticity vectors of dense particle 
clumps is broad (Supplementary Figure 4). This is a consequence of the SI that acts to produce 
the vertical motions needed to tilt the vorticity out of the disk midplane [2,10,17]. The 
preference for prograde rotation is established during the subsequent stage of gravitational 
collapse (because prograde clumps become gravitationally bound more often than the retrograde 
clumps). This preference is consistent, for instance, with gravitational accretion of pebbles 
onto protoplanets [22]. The clump obliquities change little after the initial collapse phase 
(Supplementary Figure 2). Thus the imprint of particle concentration by the SI, modified by 
gravitational collapse, is preserved.

To predict the expected orientation of binary orbits from the SI, we rely on the published results 
of gravitational collapse simulations [3], where the initial value of $\theta$ was shown to be a 
good proxy for the binary inclination, $i_{\rm b}$.
The clump obliquity distribution can thus be directly compared to the observed distribution 
of binary inclinations [6]. Here we focus on binaries found in the dynamically cold population 
of the classical Kuiper belt (hereafter cold classicals, or CCs, defined here as members of the 
classical main belt with heliocentric orbit inclinations $i<5^\circ$; see ref. [23] for a 
definition of the classical main belt). The CCs are thought to have formed in-situ at $a>40$ au and 
survived the epoch of planetary migration relatively unharmed [24].  They probably a relatively 
pristine record of planetesimal formation. 

A comparison of the SI-model predicted binary inclinations with observations (Supplementary Table 1) 
reveals that the two distributions are indistinguishable from each other (Figure 3). Specifically, 
we implemented the Kolmogorov-Smirnov (K-S) test by comparing the cumulative distribution functions 
of observed (20) and model (over 400) binaries. The K-S test indicates that the null hypothesis 
(i.e., the two samples are drawn from the same underlying distribution) cannot be ruled out with more 
than 13\% significance. This comparison clinches an argument in favor of the SI. Crucially, there is 
a marked $\simeq$4:1 preference for prograde orbits ($i_{\rm b}<90^\circ$). For comparison, ref. 
[7] proposed that the equal-size binaries in the Kuiper belt formed by capture during the coagulation 
growth of planetesimals. Capture in their model presumably occurred as a result of dynamical friction 
from a sea of small planetesimals (the L$^2$s mechanism in [7]) or via three-body encounters 
(the L$^3$ mechanism). These capture models can be ruled out based on the observed inclination 
distribution of binaries ([6]; Figure 3), because the L$^2$s mechanism predicts retrograde binary orbits 
with $i_{\rm b} \sim 180^\circ$, whereas the L$^3$ mechanism implies a $\sim$3:2 preference for 
retrograde orbits [25].

The SI is expected to occur over a wide range of protoplanetary disk conditions and pebble sizes 
[8], which suggests that the planetesimal formation by the SI was widespread. Previous studies 
explored the SI implications for the initial size distribution (ISD) of planetesimals (e.g., [16,26]),
which can be described by a rolling power-law function with an exponential cut-off at large sizes. 
Attempts to validate the ISD on the size distribution of asteroids and Kuiper belt objects are 
obscured by secondary processes that modified the distributions after formation (e.g., sustained 
impact fragmentation, [27]). Here, we point out that the ISD expected from the SI appears to be 
broadly consistent with the rounded profile of the absolute magnitude distribution of the CCs [28]. 

The distinctive shape of the New Horizons flyby target (486958) 2014 MU$_{69}$, which belongs to the 
CC population, provides additional constraints on the planetesimal formation process.
MU$_{69}$ is a contact binary consisting of two lenticular lobes, roughly $22\times 20 \times 7$ km 
and $14 \times 14 \times 10$ km in size, connected by a narrow neck [29]. The New Horizons team 
have been interpreting the shape as resulting from gentle gravitational collapse [29]. The CC 
binaries apparently span the full range of component separations and sizes from widely separated 
$\sim$100-km-class binaries to contact/small binaries such as MU$_{69}$. They are the key 
to understanding the protoplanetary disk conditions at $>$30 au during planetesimal formation.

\clearpage

\noindent
{\bf Corresponding author}\\
David Nesvorn\'y\\
Southwest Research Institute\\
1050 Walnut St., Suite 300\\
Boulder, Colorado 80302\\
Phone: (303) 546-0023\\
Email: davidn@boulder.swri.edu     

\noindent
{\bf Acknowledgments}\\
The work of D.N. was funded by the NASA Emerging Worlds program. R.L. acknowledges support from NASA
grant NNX16AP53H. A.N.Y. acknowledges support from NASA through grant NNX17AK59G and the NSF through 
grant 1616929. The funding sources of J.B.S. are NASA grants NNX13AI58G, NNX16AB42G, 80NSSC18K0640, 
and 80NSSC18K0597. W.M.G contribution was supported in part by NASA Keck PI Data Awards, administrated 
by the NASA Exoplanet Science Institute and in part by data analysis grants from the Space Telescope 
Science Institute (STScI), operated by the Association of Universities for Research in Astronomy, Inc., 
(AURA) under NASA contract NAS 5-26555.

\noindent
{\bf Competing Interest Statement}\\
The authors declare no competing interests.

\noindent
{\bf Author contributions}\\{}
D.N. suggested a comparison of clump obliquities with binary orbit inclinations and prepared the
manuscript for publication. R.L. ran one of the ATHENA simulations and performed data analyses 
with {\tt PLAN}. A.N.Y developed scaling relations for planetesimal mass estimates. J.B.S. ran 
two of the {\tt ATHENA} simulations. W.M.G. provided the data on trans-Neptunian binaries. 
All authors contributed to the interpretation of the results and writing of this paper.

\noindent{\bf Author information}\\
Reprints and permissions information is available at www.nature.com/reprints.\\
Correspondence and requests for materials should be addressed to davidn@boulder.swri.edu.

\clearpage 
\noindent{\bf Methods}

\noindent{\bf ATHENA code }\\[1.mm] 
Our numerical simulations use the {\sc Athena} code. 
{\sc Athena} is a second-order accurate, flux-conservative Godunov code for solving (in this 
particular application) the coupled equations of hydrodynamics and particle dynamics.  
To properly integrate the gas dynamics, we have employed the dimensionally unsplit corner 
transport upwind method [31] and the piecewise parabolic method [32]
to spatially reconstruct gas quantities in a third order accurate manner. The calculation of
numerical fluxes is done via the HLLC Riemann solver 
[33]. An in-depth description of the base {\sc Athena} algorithm, along with tests of this 
algorithm, can be found in ref. [18].

Particles in the {\sc Athena} code are treated via the super-particle approach, in which each 
super-particle is a statistical representation of a larger swarm of particles. 
The super-particles' (hereafter just `particles' for brevity) equations of motion 
are integrated via a semi-implicit drift-kick-drift method (see ref. [34]), 
and the momentum exchange between the particles and gas is handled via the triangular 
shaped cloud (TSC) scheme [35,36]. This approach maps the particle 
momentum from the particle locations to the grid-cell centers (where hydrodynamic 
quantities are handled) and conversely the gas velocity to the particle locations.  
A detailed description and tests of the particle integration algorithm can be 
found in ref. [34]. 

\noindent{\bf Shearing box}\\[1.mm] 
All of our simulations are carried out in the local, shearing box approximation.  
This approximation treats a co-rotating portion (or ``patch") of an accretion disk of 
length scale $L \ll R$, where $R$ is the distance from the central star. 
In this limit, the relevant equations can be expanded into a Cartesian basis $(x,y,z)$, 
which is defined relative to disk cylindrical coordinates as $x=(R-R_0)$, $y=R_0 \phi$, 
and $z = z$, where $R_0$ is a reference distance. 
Furthermore, appropriate source terms must be added to the equations to 
account for the non-inertial reference frame of the domain. See ref. [37] for a detailed 
description of the shearing box and refs. [34,38] for a description of the shearing box 
within the context of particle-gas calculations. 

The co-rotation frequency of the shearing box equals the Keplerian frequency at $R_0$, 
which we define as $\Omega$.  In our particular setup, we include the vertical component 
of the star's gravity (i.e., the simulations are vertically stratified). 
All of our simulations also employ the orbital advection (or FARGO) algorithm 
[39,40] to analytically integrate quantities along the Keplerian shear flow and integrate 
the perturbed (relative to Keplerian) quantities using the algorithms described above.  
This is done for both the gas and the particles and improves the speed and accuracy of 
the code.  To preserve epicyclic energy to machine precision, Crank-Nicholson differencing 
is used to integrate the non-inertial, shearing box source terms.  Finally, we assume 
an isothermal equation of state for simplicity. 

\noindent{\bf Boundary conditions}\\[1.mm] 
The boundary conditions of our domain are shearing-periodic in the radial direction, 
purely periodic in azimuth, and a modified open/outflow in the vertical 
direction.  The radial shearing periodic boundaries are standard in the shearing box 
set-up [37]. Briefly, particles/gas that exit one radial boundary enter at the 
opposite boundary but with a displacement in azimuth and a change in angular momentum 
and energy applied to account for the difference in the orbital position.  The modified 
outflow conditions in the vertical direction consists of exponentially extrapolating 
the gas density into the ghost zones, as this significantly reduces numerical artifacts
[17,38]. Initially, these boundary conditions allow for excellent maintenance of 
hydrostatic equilibrium.  However, as the system evolves, gas will necessarily be lost 
via these vertical boundaries (particularly since our domain is quite small and thus 
the gas density at the vertical boundaries is not significantly different from that 
at the mid-plane).  To ensure mass conservation, we 
renormalize the gas density in every cell at every time step by a factor that ensures 
total mass conservation.

Crucial to driving the streaming instability is the radial pressure gradient of the 
gas.  However, a global pressure gradient is inconsistent with the radial shearing-periodic 
boundaries.  To circumvent this issue, an inward radial force is applied to the particles.  
This force creates an inward drift equivalent to that which would occur as particles 
lose angular momentum to the sub-Keplerian gas in a real disk.  This method is widely 
employed in shearing box calculations of particle-gas mixtures (e.g., [34]) 
and is well validated. 

\noindent{\bf Particle self-gravity}\\[1.mm] 
In order to study the formation of planetesimals, the mutual gravitational attraction 
between particles must be included.  When activated (see below), the gravitational force 
is added to the particle equations of motion, and this force is calculated from Poisson's 
equation.  Numerically, Poisson's equation is solved as follows.  The particle mass density 
is mapped to the grid cell centers using the TSC approach.  
This density is then mapped to the nearest time at which 
the radial boundaries are purely periodic (see ref. [37] for a discussion of this mapping 
and these ``periodic points"), after which a 3D FFT is employed to transform Poisson's 
equation into Fourier space.  

The gravitational potential is then solved in Fourier space, after which it is transformed 
back to real space using a second 3D FFT and then mapped back to the original non-periodic 
frame.  As our vertical boundaries are open, a Green's function approach is used to solve 
Poisson's equation in the vertical direction (see details in [38,41]). Finally,
the gravitational force is calculated by a second-order, central finite difference of 
the gravitational potential and then mapped from the grid cell centers back to the locations 
of the particles via TSC.  Boundary conditions must also be applied to the gravitational 
potential, and these are essentially the same as for the gas quantities: shearing-periodic 
in $x$, purely periodic in $y$, and open in $z$ with the potential extrapolated into the 
ghost zones via a third-order scheme. The algorithm has been tested rigorously in ref. [38]. 

\noindent{\bf Initial conditions}\\[1.mm] 
All of our simulations have the following initial conditions. The gas is in
hydrostatic balance with vertical gravity, resulting in an initially Gaussian profile
\begin{equation}
\label{gas_profile}
\rho_g = \rho_0\,  {\rm exp}\left(\frac{-z^2}{2H^2}\right)\ ,
\end{equation} 
where $\rho_0$ is the mid-plane gas density, and $H$ is the vertical scale height 
of the gas. We set the standard gas parameters equal to unity; $\rho_0 = H = 
\Omega = \cs = 1$ ($\cs$ is the sound speed).

The particles are initially distributed with a Gaussian vertical profile, with particle 
scale height $H_p = 0.02H$, whereas in $x$ and $y$, the particles are uniformly distributed.  
Initially, all perturbed velocities (gas and particle velocities with the Keplerian shear 
subtracted) are set to zero. Finally, we introduce random noise (sampled from a uniform 
distribution) to the particle locations in all three dimensions to seed the streaming 
instability. 

We first run our numerical simulations without particle self-gravity, 
allowing the streaming instability to fully develop and produce particle clumping.  After 
the streaming instability has saturated (which is generally defined by when the maximum 
particle density is stochastically varying but statistically constant in time), we switch 
on self-gravity.  This method is done largely for numerical convenience, as particle 
self-gravity imposes a performance hit on the integration.  In other works [38,42], we 
have verified that the start time of self-gravity does not largely influence the final 
properties of planetesimals. 

\noindent{\bf Simulation parameters}\\[1.mm] 
In general, streaming-induced planetesimal formation calculations can be characterized 
via four dimensionless quantities [16,42]: the dimensionless stopping time,
\begin{equation}
 \tau = t_{\rm stop} \Omega\ ,
\end{equation}
which is the ratio of the dimensional stopping time $t_{\rm stop}$ (i.e., the timescale 
over which particle momentum relative to the gas motion decreases by $1/e$) to the dynamical time 
$\Omega^{-1}$ and characterizes the gas-particle interaction (for a single particle species 
in this case), the particle concentration,
\begin{equation}
 Z = \frac{\Sigma_p}{\Sigma_g}\ ,
\end{equation}  
which is the ratio of the particle mass surface density $\Sigma_p$ to the gas surface 
density $\Sigma_g$ (note that this parameter is related to, but not the same as 
'metallicity'), and a radial pressure gradient parameter that accounts for the 
irradiation from the central star and thus the sub-Keplerian rotating gas in real disks,
\begin{equation}
\label{headwind}
\Pi = \eta v_{\rm K}/\cs \ ,
\end{equation}
\noindent
where $v_{\rm K}$ is the Keplerian velocity and  $\eta$ is related to the degree to which 
gas orbital velocities are sub-Keplerian: 
%$v_\phi = v_{\rm K} \sqrt{1-\eta}$.
$v_{\phi} = v_{\rm K} (1-\eta)$ in the shearing box approximation.

When particle self-gravity is activated, an additional parameter comes into play, which
is the relative strength of self-gravity to tidal shear.  We quantify this relative 
strength as
\begin{equation}
\label{Gtilde}
\tilde{G} = \frac{4\pi G\rho_0}{\Omega^2}. 
\end{equation}
This parameter can be related to the Toomre $Q$ via $\tilde{G} = (4/\sqrt{2\pi}) Q^{-1}$. 
Higher values of $\tilde{G}$ (lower values of $Q$), which occur at larger distances 
from the central star, equate to weaker tidal shear relative to gravitational binding. 

\clearpage
\noindent{\bf Simulation setup}\\[1.mm] 
We have run three simulations in total that span a range of $\tau$, $Z$ and resolution. Run B22 
has $\tau = 2$, $Z = 0.1$, and a domain size $L_x\times L_y\times L_z$ = $(0.2H)^3$ 
with resolution $ N_x\times N_y\times N_z = 512^3$. Run C203 has $\tau = 0.3$, 
$Z = 0.02$, and a domain size $L_x\times L_y\times L_z$ = $(0.2H)^3$ with 
resolution $ N_x\times N_y\times N_z = 512^3$. Finally, simulation A12 has $\tau = 2$, 
$Z = 0.1$, and a domain size $L_x\times L_y\times L_z$ = $0.1H\times0.1H\times0.2H$ with resolution 
$N_x\times N_y\times N_z = 512\times512\times1024$. All simulations 
have $\tilde{G} = 0.05$, which equates to a Toomre $Q \simeq 32$, $\Pi = 0.05$, and are initiated 
with $N_{\rm par} = 1.34\times10^8$ (run A12) or $1.54\times10^8$ particles (B22 and C203). 

The relation between particle size and $\tau$ depends on uncertain properties of the gas disk.
If, for reference, we adopt values of the MMSN from ref. [43], a cm-size particle 
at 45 au would have $\tau = 0.3$. Ref. [12] argued that CCs formed when the 
gas density at 45 au was reduced by photoevaporation. If, for example, the gas density was 
reduced ten fold, the range of $\tau$ adopted here would correspond to sub-cm-size particles. 
Alternately, ref. [42] considered a gas disk model with surface density $\Sigma(r) \propto 1/r$, 
outer radius at 500 au, and total mass $M_{\rm disk}=0.05 M_\odot$, where $M_\odot$ is the solar mass. 
This disk model would give $\tau=0.3$ for a 2.5 mm particle and $\tau=2$ for a 1.7 cm 
particle (both at 45 au).

\noindent{\bf Clump identification}\\[1.mm] 
To identify and further characterize the properties of planetesimals produced in our simulations, 
we use a newly developed clump-finding tool, PLanetesimal ANalyzer (\texttt{PLAN}, [44]).  
It is designed to work with the 3D particle output of {\sc Athena} and find gravitationally-bound 
clumps robustly and efficiently.  \texttt{PLAN}, which is written in 
C\nolinebreak\hspace{-.05em}\raisebox{.4ex}{\tiny\bf +}\nolinebreak\hspace{-.10em}\raisebox{.4ex}{\tiny\bf +} 
with OpenMP/MPI, is massively parallelized to analyze billions of particles and many 
snapshots simultaneously.  

Briefly, the workflow of our clump-finding algorithm in \texttt{PLAN} is as follows.  The 
approach is based on the halo finder \texttt{HOP} [45], which allows for fast grouping of 
physically related particles.  \texttt{PLAN} first builds a memory-efficient linear Barnes-Hut 
tree representing all the particles [46,47].  Each particle is then 
assigned a density computed from the nearest $N_{\rm den}$ particles ($N_{\rm den} = 64$ by default).  
For particles with densities higher than a threshold, $\delta_{\rm outer} = 8\rho_{0}/\tilde{G}$, 
\texttt{PLAN} chains them up towards their densest neighbors repetitively until a density 
peak is reached.  All the particle chains linked to the same density peak are combined to 
create a group.  

\texttt{PLAN} then merges those groups by examining their boundaries to construct a list of 
bound clumps.  Based on the total kinematic and gravitational energies, deeply intersected 
groups are merged if bound.  However, two particle groups with a saddle point less dense 
than $\delta_{\rm saddle}=2.5\delta_{\rm outer}$ remain separated [45].  Next, \texttt{PLAN} goes 
through each group --- or raw clump --- to unbind any contamination (i.e., passing-by and 
not bound) particles and gather possibly unidentified member particles within its Hill sphere.  
After discarding those clumps with Hill radii smaller than one hydrodynamic grid cell ($\Delta x$) 
or density peaks less than $\delta_{\rm peak}=3\delta_{\rm outer}$, \texttt{PLAN} outputs the 
final list of clumps with their physical properties derived from particles.

Unlike \texttt{HOP}, \texttt{PLAN} does not further diagnose sub-structures within clumps.  
Most clumps in our simulations are highly-concentrated, where particles often collapse into 
regions much smaller than the Hill radius but comparable to or smaller than $\Delta x$, 
invalidating the search for sub-cell structures.  The exact internal architecture of 
self-bound clumps is beyond the scope of this work.  Nevertheless, \texttt{PLAN} does a robust 
job of identifying clumps as seen in this work (Figure 1) and other recent work [42].  

\noindent{\bf Clump properties}\\[1.mm] 
In order to determine the inclination of bound clumps, we use \texttt{PLAN} to calculate 
their angular momenta in the inertial frame by summing up the contribution from each particle 
[48,49].  Assuming that the $i$-th particle of a clump is located at $(x_i, y_i, z_i)$ relative 
to the center-of-mass of the clump, the inertial frame angular momentum $\bmath{J} = (J_x, J_y, J_z)$ 
can be written as
\begin{align}
J_x &= \sum\limits_i m_i (y_i \dot{z_i} - z_i \dot{y_i} - \Omega x_i z_i), \\
J_y &= \sum\limits_i m_i (z_i \dot{x_i} - x_i \dot{z_i} - \Omega y_i z_i), \\
J_z &= \sum\limits_i m_i [x_i \dot{y_i} - y_i \dot{x_i} - \Omega (x_i^2 + y_i^2)],
\end{align}
where $m_i$ is the particle mass. The obliquity of such a clump (i.e., the colatitude of 
$\bmath{J}$) is then calculated from 
\begin{equation}
\theta = \langle\bmath{J}, \hat{z}\rangle = \cos^{-1}\left(\frac{\bmath{J}\cdot\hat{z}}{|\bmath{J}|}\right)\ .
\end{equation}

\clearpage

\noindent{\bf Code availability}\\{}
The {\tt ATHENA} code is available on GitHub (https://github.com/PrincetonUniversity/Athena-Cversion). 
The {\tt PLAN} code is available on Zenodo (see ref. [44]).

\noindent{\bf Data availability}\\
The data that support the plots within this paper and other findings of this study are available 
from the corresponding author upon reasonable request.

\clearpage
\begin{figure}
\epsscale{1.0}
\plotone{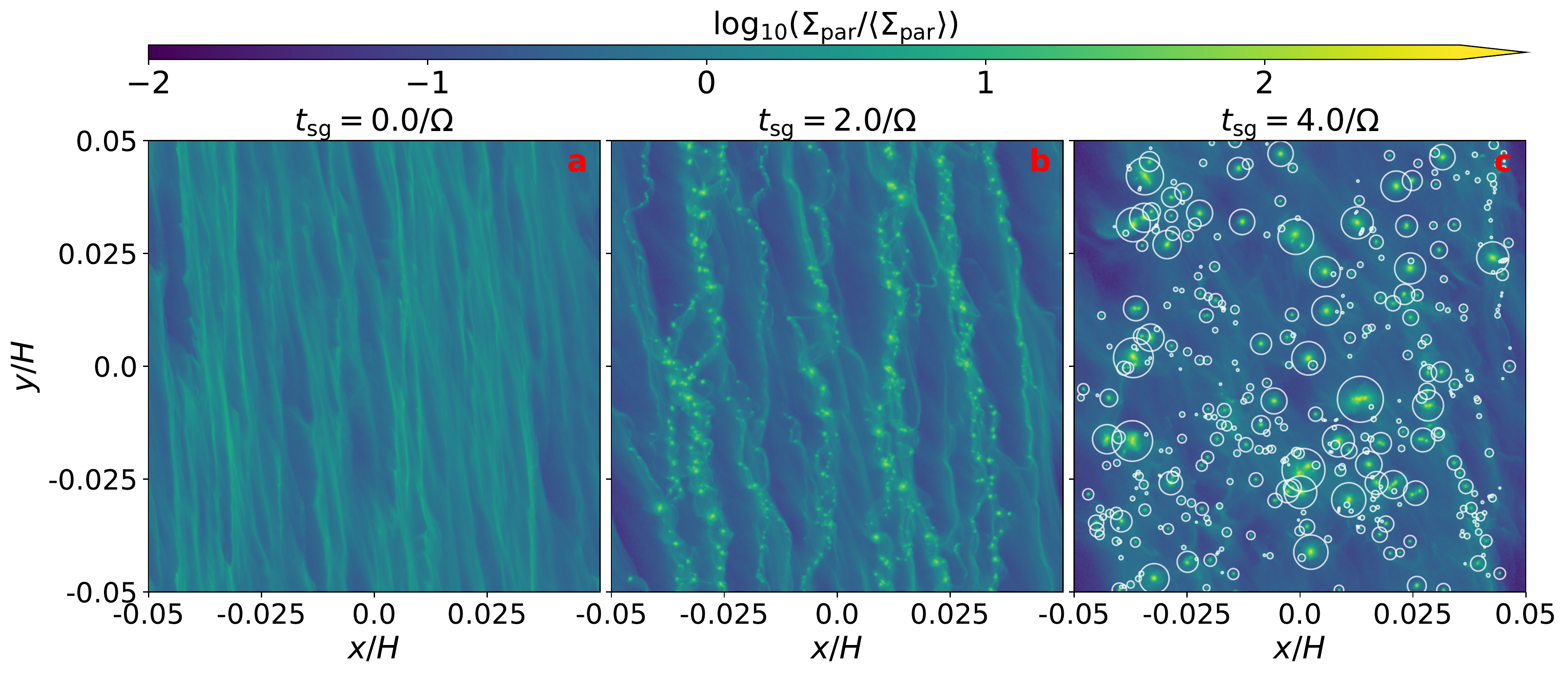}
\caption{Three snapshots from our 3D simulation of the streaming instability where the nonlinear 
particle clumping triggers gravitational collapse into planetesimals. The plots show the 
vertically-integrated density of solids ($\Sigma_{\rm par}$), projected on the disk plane, relative 
to the initially uniform surface density ($\langle\Sigma_{\rm par}\rangle$). 
The $x$ and $y$ coordinates show the shearing box dimensions 
in units of the gas disk scale height, $H$ (the Sun is to the left, the orbital velocity vector points 
up). Time $t$ increases from left to right as labeled ($t_{\rm sg} = t - t_0$; particle self-gravity was 
switched on at $t_0=36/\Omega$). Azimuthal filaments have already formed at $t_{\rm sg}=0$ (panel a). 
In b and c, the filaments fragment into gravitationally bound clumps. The circles in panel c depict 
the Hill spheres of clumps that were identified by the {\tt PLAN} algorithm. The full evolution is 
shown in the Supplementary Video.}
\label{f1}
\end{figure}

\clearpage
\begin{figure}
%/d1/Binary_lhotse1_d2/gr.ratio_f1
\epsscale{0.8}
\plotone{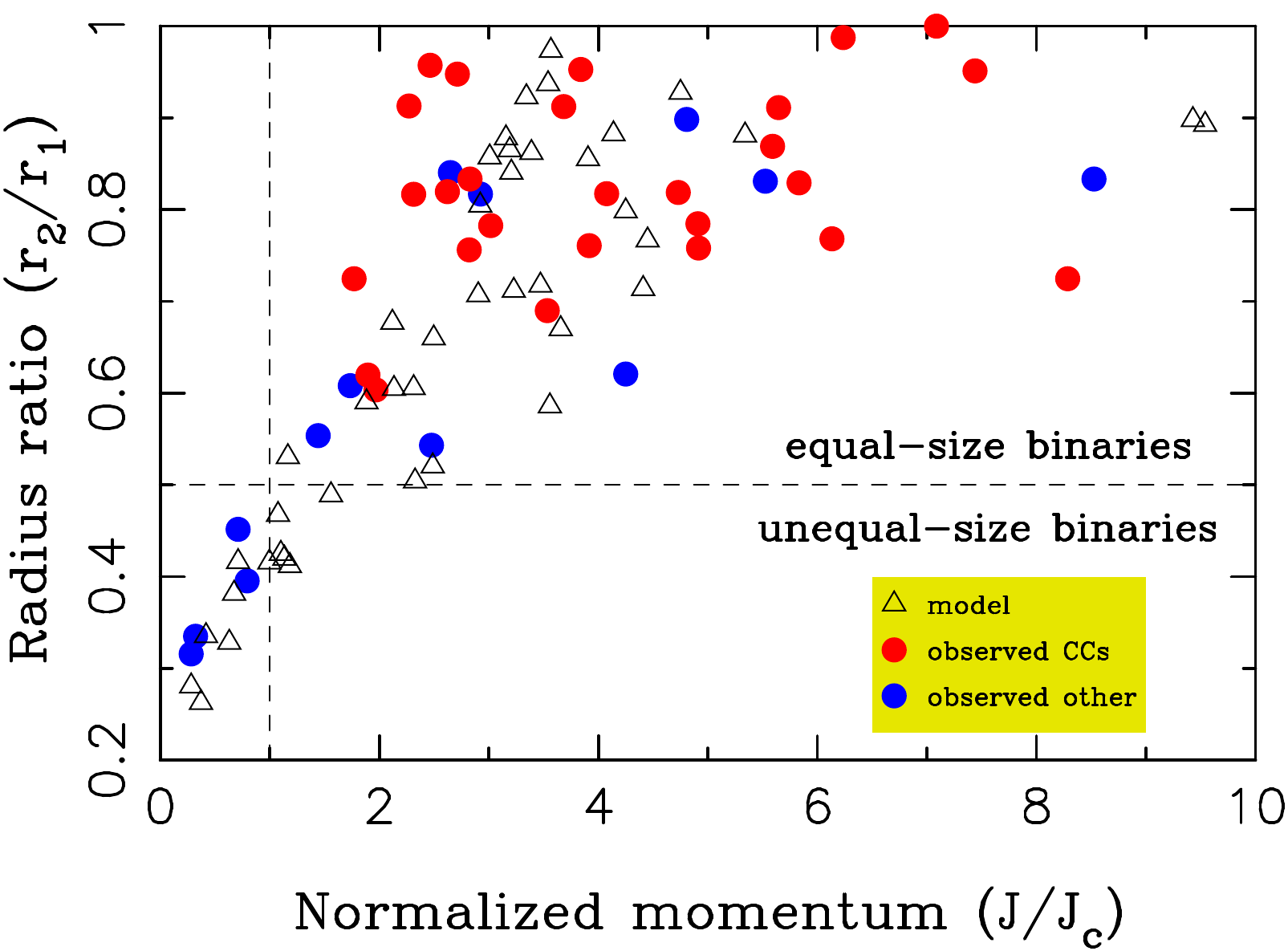}
\caption{The matching properties of model (triangles) and observed (red and blue dots) binary planetesimals. 
The model results were taken from new gravitational collapse simulations similar to those reported in 
ref. [3]. As indicated by our SI simulations, the gravitationally bound clumps have vigorous rotation. Here, 
we therefore plot cases with the initial rotation $\omega \geq 0.5 V_{\rm circ}/R$, where $V_{\rm circ}$ is the 
circular speed of a particle at the clump radius $R$. The known trans-Neptunian binaries were separated into 
binaries found among the cold classicals (as defined by the heliocentric orbit inclination $i<5^\circ$; red dots) 
and everything else (blue dots). Note that some of the known unequal-size binaries can be impact generated. 
The source of the observational data is the Minor Planet Center database.}
\label{f2}
\end{figure}

\clearpage
\begin{figure}
\epsscale{0.8}
\plotone{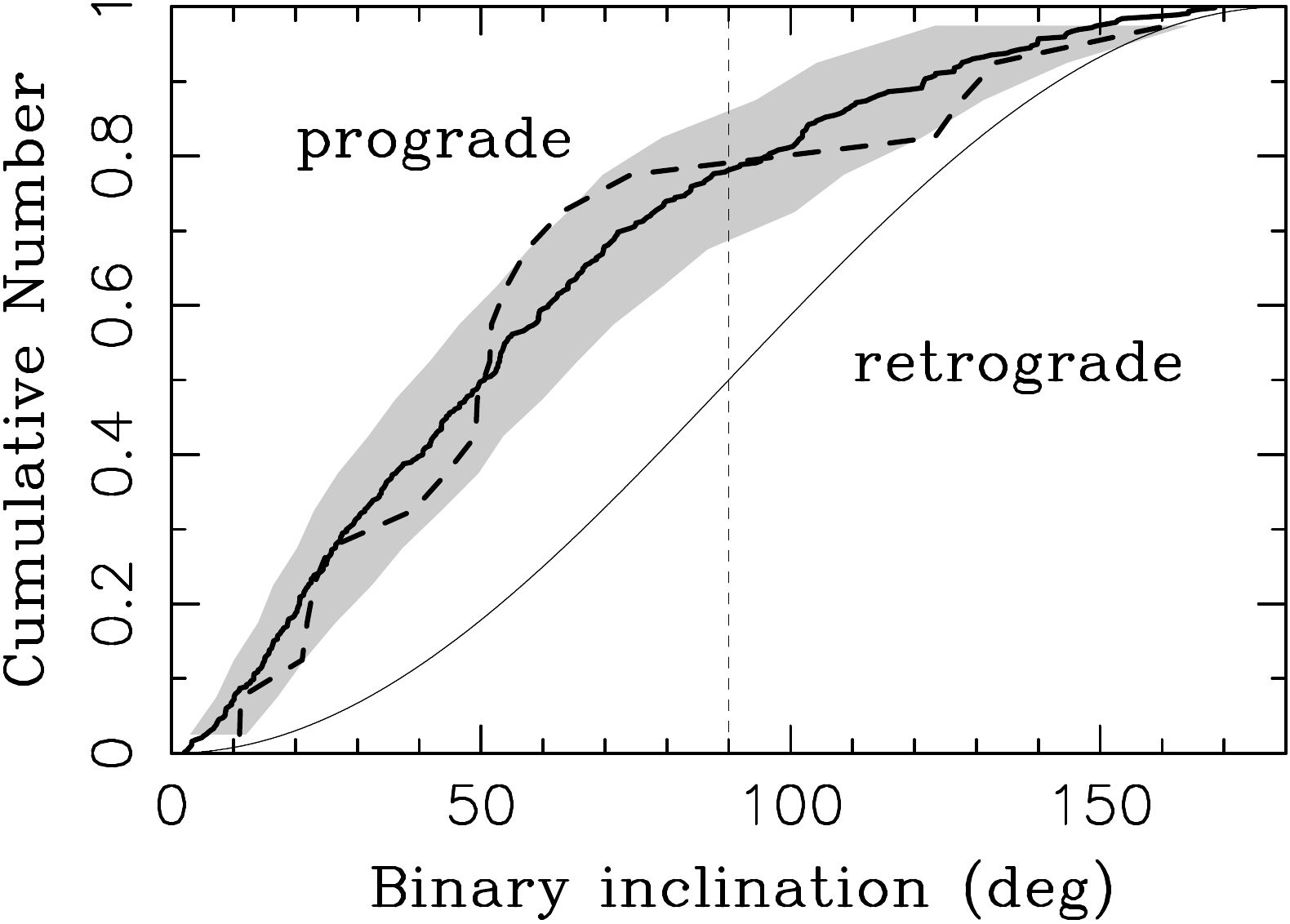}
\caption{The inclination distribution of binary orbits obtained in the SI model (bold solid line) 
matches observations of trans-Neptunian binaries (bold dashed line). Here we plot the results 
of our run A12 at time $t_{\rm sg} = 4 / \Omega$. The observed inclination distribution shown 
here includes all CC binaries from ref. [6] with $a_{\rm b}<0.1$~$R_{\rm Hill}$, where $R_{\rm Hill}$ 
is the Hill radius. Binaries with $a_{\rm b}>0.1$~$R_{\rm Hill}$ are not plotted to limit
potential effects of the solar gravity, which may alter wide binary orbits with 
$i_{\rm b} \sim 90^\circ$ (see Supplementary Materials), and gravitational encounters with large 
bodies [30]. We note that it makes no significant difference whether the wide 
binaries are included or excluded, because a great majority of known binaries has 
$a_{\rm b}<0.1$~$R_{\rm Hill}$ (Supplementary Table 1). The shaded area is a 68\% envelope 
of the model results when a sample of 20 model orbits is randomly drawn (i.e., sample size equal 
to that of real binaries with known inclinations; see Supplementary Figure 5 for a similar 
comparison with 100 model orbits). The binary orbits are predominantly prograde with $\simeq$80\% 
having $i_{\rm b}<90^\circ$. The cumulative distribution of inclinations for randomly oriented binary 
orbits (thin solid line) is plotted for reference. }
\label{f3}
\end{figure}

\end{document}